\documentclass[12pt,aps]{revtex4}

\usepackage{graphicx} 

\usepackage{amsfonts}

\begin{document}

\renewcommand{\figurename}{Fig.}

\title{Quantum mechanical analogy and supersymmetry of electromagnetic wave modes in planar waveguides}

\author{H. P. Laba$^1$, V. M Tkachuk$^2$\footnote{voltkachuk@gmail.com}
}
\address{ $^1$Lviv Polytechnic National Univerity,
Department of Telecomunications,
12, S. Bandery Street, 79013, Ukraine, \\
$^2$Department for Theoretical Physics,
Ivan Franko National University of Lviv,
12, Drahomanov Street, Lviv, 79005, Ukraine.}

\begin{abstract}
We study an analogue
of the equations describing TE and TM modes in a planar waveguide with an arbitrary
continuous dependence of the electric permittivity and magnetic permeability on coordinates
with the stationary Schr\"odinger equation.
The effective potential energies involved in the Schr\"odinger equation for TE and TM modes are found. In general,
the effective potential energies for TE and TM modes are
different but in the limit of a weak dependence of the permittivity and permeability on coordinates they coincide.
In the case when the product of a position-dependent permittivity and permeability is constant,
it means that the refractive index is constant, we find that the TE and TM modes are described by the supersymmtric quantum mechanics.
\end{abstract}

\maketitle

\section{Introduction}
The analogy between wave optics and quantum mechanics and between geometric optics and classical mechanics, respectively,
is well known
(see, for instance, \cite{Black85,Evans93,Tzanakis98,Manko08} and references therein).
This analogy is not only interesting on its own right but is also important from the practical point of view.
The theoretical and experimental methods developed in one of the fields can be successfully applied in the other.
The quantum-mechanical methods, such as factorization method or sypersymmetry, variational method, WKB method and other
can be used to investigate the propagation of light in waveguides. Vice versa, the propagation of light in waveguides
can be used for experimental verification of many quantum-mechanical effects, such as quantum tunneling and time of tunneling,
Bloch oscillations in periodic structures, quantum chaos. The reviews on this subject can be found in
\cite{Longhi02,Longhi03,Orni07,Longhi09}.
Note also recent papers
\cite{Kulkarni12,Strelow12},
where the quantum-mechanical analogy was used for investigation of light confinement in
waveguide structures and in microtube bottle resonators, respectively.
Note that in \cite{Miri13}
the supersymmetric quantum mechanics was applied for investigation of complex optical potentials with real
spectra. In \cite{Miri13PRL} the authors show that supersymmetry (SUSY) can provide a versatile platform in synthesizing a new class of optical structures with desired properties and functionalities.

The aim of this paper is to obtain a quantum-mechanical analogy of equations describing TE and TM modes in planar waveguides
with an arbitrary dependence of the permittivity as well as permeability on coordinates.
Note that previously the reduction of equations describing waveguide modes to the Schr\"odinger equation was used mostly for waveguides with permeability equal to unity
(see for instance \cite{Love79,Adams81,Laba13}).
The present paper is organized as follows.
In Section 2 we write the equations describing TE and TM modes in planar waveguides with an arbitrary dependence of the permittivity and permeability on coordinates.
In Section 3 we show that these equations can be
reduced to the Schr\"odinger equation.
In Section 4, we show that in the case when the position-dependent permittivity is inverse to permeability the TE and TM modes possesses
the supersymmetric properties.
Note that supersymmetry in our case appears from another reason than in \cite{Miri13,Miri13PRL} and is the result of a special tuning of permittivity and permeability, which leads to the supersymmetry combining TE and TM modes. The authors of \cite{Miri13,Miri13PRL} consider the supersymmetry combining TE modes in optical waveguides with two different
refractive indexes that form a superpartner pair.
Finally, in Section 5 we conclude the results.

\section{Propagation of electromagnetic waves in planar waveguide}
In this section we mention the equations describing the propagation of electromagnetic field
in a planar waveguide (see, for instance, \cite{Unger77}).
Let the planar waveguide is parallel to the plane $y-z$ with permittivity
$\varepsilon=\varepsilon(x)$ and permeability $\mu =\mu (x)$, which are continuous functions of $x$ (Fig. 1).
The refractive index in this case depends on the coordinate $x$, namely $n=\sqrt{\varepsilon\mu}=n(x)$.
Such a waveguide is called the gradient waveguide.

\begin{figure}
\centerline{
\includegraphics[scale=0.6]{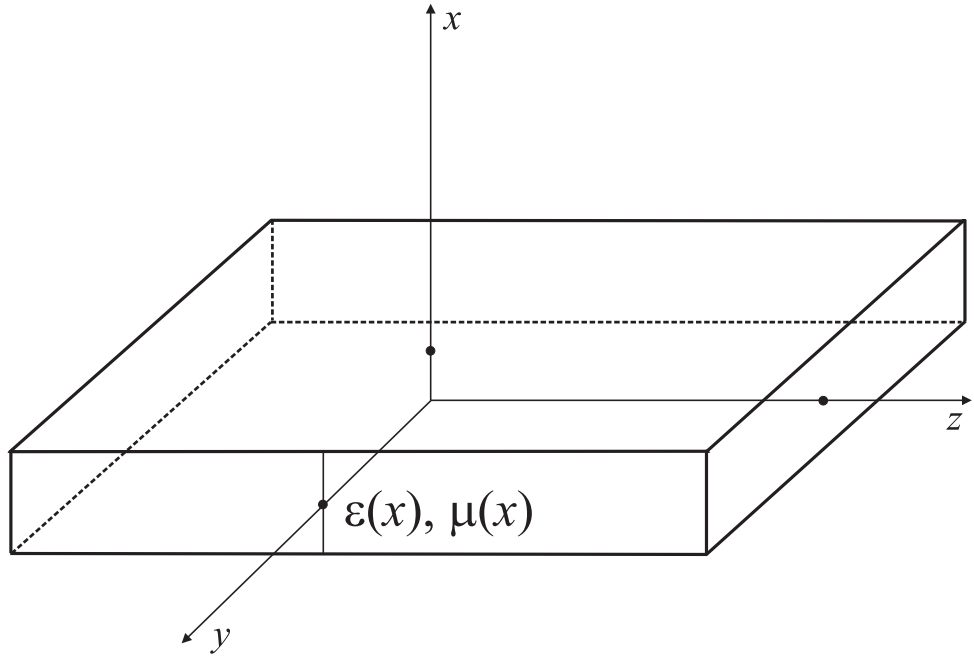}}
\caption{Planar waveguide is parallel to the $y-z$ plane. Electric permittivity and magnetic permeability vary along $x$.
Electromagnetic field propagates along $z$.}
\end{figure}

For the description of the propagation of the electromagnetic field in the planar waveguide we start from the Maxwell equations
\begin{eqnarray}\label{Max1}
{1\over c}{\partial{\bf D}\over \partial t}={\rm rot}{\bf H},
\\ {1\over c}{\partial{\bf B}\over \partial t}=-{\rm rot}{\bf E}, \label{Max2}\\
{\rm div}{\bf D}=0, \ \ {\rm div}{\bf B}=0, \label{div}
\end{eqnarray}
where ${\bf D}=\varepsilon {\bf E}$ and ${\bf B}=\mu {\bf H}$.
The solution for the electromagnetic field, which propagates in the planar waveguide along
$z$, can be found in the form
\begin{eqnarray}\label{sol0}
{\bf E}={\bf E}(x)e^{i(\beta z-\omega t)}, \ \ {\bf H}={\bf H}(x)e^{i(\beta z-\omega t)},
\end{eqnarray}
where $\beta$ is the longitudinal wavevector, which is called the propagation constant.
Substituting (\ref{sol0}) into
(\ref{Max1}) and (\ref{Max2}) we find
two independent sets of equations.
One of them contains $E_y$, $H_x$, $H_z$, which form transverse electric (TE) modes.
The second one
contains $E_x$, $E_z$, $H_y$ and form transverse magnetic (TM) modes (see, for instance, \cite{Unger77}).
One can verify that for TE and TM modes equation (\ref{div}) is satisfied.

Note that the electric field in TE mode is directed along axis $y$ and is perpendicular to the direction of propagation of the wave, the magnetic field contains also the longitudinal component. For the TM mode the situation is opposite.
Namely, the magnetic field is directed along $y$; electric field contains also longitudinal component.

The set of equations for TE modes can be reduced to the second order equation for  $E_y$ and
similarly, the set of equations for TM modes can be reduced to
the second order equation for $H_y$ respectively
\begin{eqnarray}\label{EqEy1}
-\mu(x){\partial\over\partial x }{1\over \mu(x)}{\partial\over\partial x }E_y-\varepsilon(x)\mu(x){\omega^2\over c^2}E_y+{\beta^2}E_y=0, \\
\label{EqHy1}
-\varepsilon(x){\partial\over\partial x }{1\over \varepsilon(x)}{\partial\over\partial x }H_y-\varepsilon(x)\mu(x){\omega^2\over c^2}H_y+{\beta^2}H_y=0.
\end{eqnarray}
Note that these equations are equivalent with respect to exchanging $E_y$  and $H_y$, $\varepsilon$ and $\mu$.

\section{Quantum-mechanical analogy TE and TM modes}
At the beginning of this section we note that in the case of slow dependence of permittivity and permeability
on position we can neglect $\varepsilon'(x)$ and $\mu'(x)$. Then in the first terms of equations (\ref{EqEy1}) and (\ref{EqHy1}) $\mu(x)$ and $\varepsilon(x)$ can be canceled and as a result these equations are equivalent to the Schr\"{o}dinger equation
\begin{eqnarray}\label{Schr}
-{1\over 2m}{\partial^2 \psi\over\partial x^2} +V(x)\psi={\cal E}\psi,
\end{eqnarray}
where $\hbar=1$, mass $2m=1$,  energy ${\cal E}=-\beta^2$, wave function
$\psi=E_y$ and potential energy for TE and TM modes are equivalent
\begin{eqnarray}\label{VTE}
V(x)=V_{\rm TE}(x)=V_{\rm TM}(x)=-\varepsilon(x)\mu(x){\omega^2\over c^2}.
\end{eqnarray}

Now let us consider arbitrary dependence of permittivity and permeability on position.
Following \cite{Love79}
equation (\ref{EqEy1}) can be reduced to the stationary Schr\"{o}dinger equation with some effective potential energy. It is obviously that the same can be done with equation (\ref{EqHy1}).
Introducing $E_y=A\sqrt{\mu(x)}\psi_{E}$ (where $A$ is a constant) for a new function $\psi_E(x)$ we obtain the equation
\begin{eqnarray}\label{SchrTE}
-{\partial^2 \psi_E\over \partial x^2}+\left(-\varepsilon(x)\mu(x){\omega^2\over c^2}+\left({1\over2}{\mu'(x)\over \mu(x)}\right)^2-\left({1\over2}{\mu'(x)\over \mu(x)}\right)'\right)\psi_E+\beta^2\psi_E=0.
\end{eqnarray}
This equation for TE modes is equivalent to the Schr\"{o}dinger one (\ref{Schr}),
where mass $2m=1$, $-\beta^2$ is the eigenvalue and potential energy reads
\begin{eqnarray}\label{VTE}
V(x)=V_{\rm TE}(x)=-\varepsilon(x)\mu(x){\omega^2\over c^2}+\left({1\over2}{\mu'(x)\over \mu(x)}\right)^2-\left({1\over2}{\mu'(x)\over \mu(x)}\right)'.
\end{eqnarray}

Similarly, introducing $H_y=A\sqrt{\varepsilon(x)}\psi_H$ for TM modes we obtain equation
\begin{eqnarray}\label{SchrTM}
-{\partial^2 \psi_H\over \partial x^2}+\left(-\varepsilon(x)\mu(x){\omega^2\over c^2}+\left({1\over2}{\varepsilon'(x)\over \varepsilon(x)}\right)^2-\left({1\over2}{\varepsilon'(x)\over \varepsilon(x)}\right)'\right)\psi_H+\beta^2\psi_H=0,
\end{eqnarray}
which is equivalent to the Schr\"{o}dinger one with potential energy
\begin{eqnarray}\label{VTM}
V(x)=V_{\rm TM}(x)=-\varepsilon(x)\mu(x){\omega^2\over c^2}+\left({1\over2}{\varepsilon'(x)\over \varepsilon(x)}\right)^2-\left({1\over2}{\varepsilon'(x)\over \varepsilon(x)}\right)'.
\end{eqnarray}
Thus, equations (\ref{SchrTE}) and  (\ref{SchrTM}) represent the quantum-mechanical analogy of
TE and TM modes in a planar waveguide with an arbitrary dependence of permittivity and permeability on position.

\section{SUSY of TE and TM modes in planar waveguides}

In this section we consider a special case when the refractive index
does not depend on position
\begin{eqnarray}\label{amconst}
\varepsilon(x)\mu(x)=n_0^2={\rm const}.
\end{eqnarray}
Note that in this case we have a very interesting situation. Namely, although the refractive index is constant, the
effective potential energies for TE and TM modes are not constant. Moreover, we will show that
the considered waveguide in this case exhibits SUSY properties and
TE and TM modes are related by the SUSY transformation.
A review and introduction to SUSY quantum mechanics can be found in \cite{Cooper95,Gang11}.

In case (\ref{amconst}) equations for TE and TM modes (\ref{SchrTE}) and  (\ref{SchrTM}) can be written in the form
of the supersymmetric quantum mechanics
\begin{eqnarray}\label{Schr1}
-{\partial^2 \psi_E\over \partial x^2}+\left(W^2+W'\right)\psi_E={\cal E}\psi_E.\\
-{\partial^2 \psi_H\over \partial x^2}+\left(W^2-W'\right)\psi_H={\cal E}\psi_H, \label{Schr2}
\end{eqnarray}
where the eigenvalue
\begin{eqnarray}
{\cal E}=n_0^2{\omega^2\over c^2}-\beta^2
\end{eqnarray}
and
\begin{eqnarray}
W(x)={1\over2}{\varepsilon'(x)\over \varepsilon(x)}=-{1\over2}{\mu'(x)\over \mu(x)}
\end{eqnarray}
is called superpotential.

Hamiltonians in (\ref{Schr1}) and (\ref{Schr2}) corresponding to the TE an TM modes can be factorized as follows
\begin{eqnarray}
H_{\rm TE}=-{\partial^2 \over \partial x^2}+W^2+W'=B^-B^+,\\
H_{\rm TM}=-{\partial^2 \over \partial x^2}+W^2-W'=B^+B^-,
\end{eqnarray}
where
\begin{eqnarray}
B^-={\partial \over \partial x}+W(x), \ \ B^+={\partial \over \partial x}-W(x).
\end{eqnarray}
These two Hamiltonians can be combined in one supersymmetric Hamiltonian
\begin{eqnarray}\label{Hsusy}
H=\left(
  \begin{array}{cc}
    H_{\rm TE} & 0 \\
    0 & H_{\rm TM} \\
  \end{array}
\right)=
\left(
  \begin{array}{cc}
    B^-B^+ & 0 \\
    0 & B^+B^- \\
  \end{array}
\right),
\end{eqnarray}
which acts on the two-component wave function
\begin{eqnarray}
\psi=\left(
  \begin{array}{c}
    \psi_E \\
    \psi_H \\
  \end{array}
\right).
\end{eqnarray}
Hamiltonian (\ref{Hsusy}) can be written as follows
\begin{eqnarray}
H=Q_1^2=Q_2^2,
\end{eqnarray}
where supercharges are
\begin{eqnarray}
Q_1=\left(
      \begin{array}{cc}
        0 & B^- \\
        B^+ & 0 \\
      \end{array}
    \right)=B^-\sigma^+ +B^+\sigma^-,
\end{eqnarray}
and
\begin{eqnarray}
Q_2=i\left(
      \begin{array}{cc}
        0 & B^- \\
        -B^+ & 0 \\
      \end{array}
    \right)=i(B^-\sigma^+ -B^+\sigma^-).
\end{eqnarray}
One can find that these operators satisfy the permutation relations
\begin{eqnarray}
\{Q_i,Q_j\}=2H\delta_{ij}, \ \ [Q_i,H]=0,
\end{eqnarray}
where $i,j=1,2$, $\{a,b\}$ is anticommutator and $[a,b]$ is commutator.
These permutation relations present SUSY algebra with two supercharges.
An important consequence of SUSY is the two-fold degeneracy of all nonzero eigenvalues. This statement is equivalent to the following.
Namely, the Hamiltonians $H_{\rm TE}$ and $H_{\rm TM}$ have the same nonzero eigenvalues.
Zero eigenvalue exists only for one of the Hamiltonian. Note that when zero eigenvalue exists then SUSY is called exact SUSY.
For details see review \cite{Cooper95}.

Let us consider the eigenvalue equation for TM and TE modes, which after factorization we write in the form
\begin{eqnarray}
B^-B^+\psi_E={\cal E}\psi_E. \\
B^+B^-\psi_H={\cal E}\psi_H.
\end{eqnarray}
Applying the operators $B^+$ and $B^-$ to the first and second equation respectively we find
\begin{eqnarray}
B^+\psi_E=\sqrt{\cal E}\psi_H, \ \ B^-\psi_H=\sqrt{\cal E}\psi_E.
\end{eqnarray}
These transformations give the relations between TE and TM modes in the case of supersymmetry.

Now let us consider zero eigenvalue ${\cal E}=0$, when $\beta^2=n_0^2\omega^2/c^2$. The eigenvalue equations for this case is reduced to
\begin{eqnarray}
B^+\psi_E=0, \\
B^-\psi_H=0.
\end{eqnarray}
These are first-order equations, which can be easily solved:
\begin{eqnarray}\label{psiE}
\psi_E=Ae^{\int W(x)dx}=A_E\sqrt{\varepsilon(x)}={A_En_0\over\sqrt{\mu(x)}},\\  \label{psiH}
\psi_H=Ae^{-\int W(x)dx}={A_H\over\sqrt{\varepsilon(x)}}={A_H\over n_0}\sqrt{\mu(x)},
\end{eqnarray}
where $A_H$, $A_E$ are the normalization constants.
One can see that depending on $\varepsilon(x)$ the localized solution with zero boundary condition
exists only for one of the functions $\psi_H$ or $\psi_E$ or does not exist for both functions.
In the first case we have exact SUSY and in the second one the SUSY is broken.

Let us consider an explicit example with
$\varepsilon=n_0^2e^{\alpha x^2}$ and  $\mu=e^{-\alpha x^2}$, where $\alpha>0$.
The superpotential in this case reads
\begin{eqnarray}
W=\alpha x.
\end{eqnarray}
The Hamiltonians corresponding to TM and TE modes now are
\begin{eqnarray}\label{HarTE}
H_{TE}=-{\partial^2 \over \partial x^2}+\alpha^2x^2+\alpha,\\ \label{HarTM}
H_{TM}=-{\partial^2 \over \partial x^2}+\alpha^2x^2-\alpha,
\end{eqnarray}
that are Hamiltonians of harmonic oscillators.

A square-integrable solution with zero boundary conditions on infinity for ground state with ${\cal E}=0$ given by (\ref{psiE}) and (\ref{psiH})
exists for $A_E\ne 0$ and $A_H=0$. Thus for ${\cal E}=0$ which corresponds to $\beta^2=n_0^2\omega^2/c^2$ the solution is
\begin{eqnarray}
\psi_H=A_He^{-x^2/2}, \ \ \psi_E=0.
\end{eqnarray}
So, the SUSY is exact in this example.
The solutions for all eigenvalues and corresponding eigenfunctions is well known and can be found in differen ways. One of then is factorization method or SUSY method.
Eigenvalues for (\ref{HarTE}) and (\ref{HarTM}) reads respectively
\begin{eqnarray}
{\cal E}_{{\rm TE},n}=2\alpha(n+1), \\
{\cal E}_{{\rm TM},n}=2\alpha n,
\end{eqnarray}
where $n=0,1,2,\ldots$.
Thus for TE and TM modes we have
\begin{eqnarray}
\beta^2_{{\rm TE},n}=n_0^2{\omega^2\over c^2}+2\alpha(n+1), \\
\beta^2_{{\rm TM},n}=n_0^2{\omega^2\over c^2}+2\alpha n .
\end{eqnarray}
Note that in the result of the SUSY
\begin{eqnarray}
{\cal E}_{{\rm TE},n}={\cal E}_{{\rm TM},n+1}
\end{eqnarray}
and respectively the propagation coefficients for TE and TM modes are related as follows
\begin{eqnarray}
\beta^2_{{\rm TE},n}=\beta^2_{{\rm TM},n+1}.
\end{eqnarray}
As we see, all the modes except the ground one with zero energy, are two-fold degenerate, namely,
for the same propagation coefficient in a planar waveguide the TE and TM modes exist simultaneously.
At the smallest propagation coefficient $\beta=n_0^2\omega^2/c^2$, only TM mode exists for this example.

\section{Conclusions}\emph{}
In this paper
we establish an analogue
of the equations describing the TE and TM modes in a planar waveguide with an arbitrary
continuous dependence of electric permittivity and magnetic permeability on coordinates
with the stationary Schr\"odinger equation.
Namely, we show that equations for TE and TM modes in a planar waveguide can be reduced to the Schr\"odinger equation with some effective potential energy for each of the modes (\ref{VTE}) and (\ref{VTM}).
In general, the effective potential energies for TE and TM modes involved in the Schr\"odinger equation are
different but in the limit of a weak dependence of the permittivity and permeability on coordinates they coincide.

There is an interesting
case when the product of a position-dependent permittivity and permeability is constant.
Although
the refractive index in this case is constant the electromagnetic wave feels the position-dependent permittivity and permeability separately and considered structure works as a waveguide.
In this case we find that the TE and TM modes is described by the SUSY quantum mechanics.
Therefore, all properties of SUSY are transferred on the propagation of electromagnetic wave in planar waveguide.
Namely, in the case of exact SUSY there exists non-degenerate zero energy ground state.
This statement is equivalent to the following.
For the smallest propagation coefficient $\beta^2=n_0^2\omega^2/c^2$, only
one of the modes (TE or TM) exists in a planar waveguide.
The next statement of SUSY quantum mechanics tells that all nonzero energy states are two-fold degenerate.
It means that with the same propagation coefficient $\beta^2>n_0^2\omega^2/c^2$
two modes TE and TM exist
in planar waveguide, simultaneously.
To achieve SUSY case experimentally, the position-dependent permittivity must be inverse to permeability and then the refractive index is constant (see (\ref{amconst}) and example at the end of Section 4).
It seems that this very tuning is the most complicated problem for experimentalists.
We hope that recent progress with metamaterials (see, for instance, \cite{Wenchan10}) gives a possibility to overcome this problem and realize SUSY in planar waveguides on the experimental level.
It will allow to see experimentally all SUSY properties in planar waveguides.
One of them is the two fold degeneracy that meaning that TE and TM modes exist with the same propagation coefficient.
On the quantum level it means that for a fixed propagation coefficient we have a two-dimensional quantum space. It probably can be used for a realization of quantum bit which is a basic element in quantum information.

\end{document}